\documentstyle[psfig,rotate]{l-aa}
\input{epsf}
\newcommand{\refer}{\bibitem{}}

\newcommand{\vecv}{{\bf v}}
\newcommand{\vecb}{{\bf B}}

\begin{document}
   \thesaurus{06         % A&A Section 6: Form. struct. and evolut. of stars
              (02.13.2;
               02.19.1;
               09.03.1;  
               09.10.1;
	       09.13.2;
               13.09.4;)} 

\title{The  formation of C-shocks: structure and signatures}

\author{Michael D. Smith$^1$ \& Mordecai-Mark Mac Low$^2$}
\offprints{M.D. Smith}

\institute{$^1$Astronomisches Institut der Universit\"at W\"urzburg, Am Hubland,
D-97074 W\"urzburg, Germany\\
$^2$Max-Planck-Institut f\"ur Astronomie, K\"onigstuhl 17, D-69117
Heidelberg, Germany\\
Internet: smith@astro.uni-wuerzburg.de,  mordecai@mpia-hd.mpg.de}

   \maketitle

   \markboth{Smith \& Mac Low}{Formation of C-shocks}

\begin{abstract}
Shock waves in molecular clouds should evolve into continuous or C-type
structures due to the
magnetic field and ion-neutral friction. We here determine whether and
how this is achieved 
through plane-parallel numerical simulations using an extended version
of ZEUS. 
We first describe and test the adapted code against analytical
results, laying the necessary foundations for subsequent 
works on supersonic ambipolar diffusion, including C-type jets and
shock instability. 

The evolution away from jump shocks 
toward the numerous steady C-shock sub-types  is then investigated.
The evolution passes through
four stages, which possess distinctive observational properties.
The time scales and length scales cover broad ranges. Specific results
are included for shock types including switch, absorber, neutralised, oblique, transverse
and intermediate.  Only intermediate
Type II shocks and `slow shocks', including switch-off shocks, remain as J-type under the low ion levels assumed.
Other shocks transform via a steadily growing  neutral precursor to a
diminishing jump.
For neutralised shocks, this takes the form of an extended long-lived ramp.

Molecular hydrogen emission signatures are presented. After the jump speed has dropped to under 25\,km\,s$^{-1}$, a non-dissociative jump section can 
dominate the spectra for a long period. This produces a high-excitation
spectrum. Once the jump has further weakened, to $<$\,8\,km\,s$^{-1}$,  the fully developed  ion front
is responsible for brisk progress towards a constant C-type
excitation. The time scale for emission-line variations is $\sim (6
/n_i$)\,yr, where n$_i$ is the pre-shock ion number density. 
\end{abstract}


%%%%%%%%%%%%%%%%%%%%%%%%%%%%%%%%%%%%%%%%%%%%%%%%%%%%%%%%%%%%%%%%%%%%%%%%%%%%
%%%%%%%%%%%%%%%%%%%%%%%%%%%%%%%%%%%%%%%%%%%%%%%%%%%%%%%%%%%%%%%%%%%%%%%%%%%%
\section{Introduction}
%%%%%%%%%%%%%%%%%%%%%%%%%%%%%%%%%%%%%%%%%%%%%%%%%%%%%%%%%%%%%%%%%%%%%%%%%%%%
%%%%%%%%%%%%%%%%%%%%%%%%%%%%%%%%%%%%%%%%%%%%%%%%%%%%%%%%%%%%%%%%%%%%%%%%%%%%

C-type shocks are frequently invoked to interpret the signatures of accelerated
and excited gas in molecular clouds. These magnetically-mediated, two-fluid shocks are able to explain the survival of the molecules in shocks with velocities
exceeding 20\,km\,s$^{-1}$
as well as the low excitation of observed spectra (e.g. the strong atomic fine-structure lines of oxygen  and the low values of the 2-1 S(1)/1-0 S(1) ratio of molecular hydrogen). However, several basic questions on the
existence of C-shocks remain to be answered.
 Under what conditions do steady-state C-shocks form? Is there sufficient time for the various C-shock
configurations to be realized?  Alternatively, how can we recognize an incomplete or proto-C-shock? Here,
as part of a wider program to investigate supersonic ambipolar diffusion,
we develop a numerical model and apply it to study the time-dependence of planar C-shock configurations.

A theory for steady C-shocks in molecular gas was constructed by  Draine (1980)
and applied by  Draine, Roberge \& Dalgarno (1983). The combination of
low ion fraction, strong cooling, and significant (but not high) magnetic
field results in a shock in which ion-neutral drag provides the
viscosity, cooling keeps the gas supersonic and the field via the ions
provides an extended  cushioning layer which inhibits molecular dissociation (Draine \& McKee 1993). Steady solutions for transverse-field C-shocks were further analysed by  Chernoff (1987), Roberge \& Draine (1990) and
Smith \& Brand (1990a). Smith, Brand \& Moorhouse (1991) looked at the high-field
`shock absorbers'. Wardle \& Draine (1987) \& Smith (1993a,b) presented a theory for oblique-field C-shocks, including the parallel-field C-switches. 
Flower, Pineau des For\^ets \& Hartquist (1985) considered the chemical 
aspects of steady C-shocks, work which has developed into a theory for
C-shock chemistry (see Flower et al 1996). Predictions for steady, planar C-shocks were presented, in addition, by Smith \& Brand (1990b), Smith (1991, 1995) and Kaufman \& Neufeld (1996a,b).

Are these steady-state predictions relevant? There are two other possibilities:
instability and evolution. Instability has been investigated by Wardle
(1990, 1991a,b), To\'th (1994,1995) and now by Stone (1997) and Mac Low
\& Smith (1997a,b). Conditions for stability are either (i) neutral
Alfven numbers $M_a$ = v$_s$/v$_a$ $< 5$ ,where v$_s$ and v$_a$ = B /$(4\pi\rho_n)^{1/2}$ are the shock and Alfv\'en speeds, respectively,
or (ii) rapid recombination so that the ion fraction is fixed locally
rather than
via advection.
However, it is not  clear that the steady state is approached even in the
absence of instability.
This evolutionary question, which has not been examined before (Pineau des For\^ets 1997), is the focus of this paper. Furthermore, we  predict the speed and
character of changes in emission line strengths that identify
a proto-C-shock, and distinguish these changes from those produced in
an unstable C-shock (as evaluated by Neufeld \& Stone (1997) and Mac Low \& Smith (1997b)).

Our attempt to evolve flow patterns from a sharply discontinuous J-shock
to a C-shock under constant ionization conditions may not exactly
correspond to a physical situation. Rather, we use these examples to envisage
how a flow accomodates to changing conditions such as those caused by
the start of a stellar outflow, the impact of a jet on a cloud, or a
cloud-cloud collision.  Our primary goal here is to
obtain a deeper understanding of how molecular shocks behave
dynamically.  We do use an alternative
initial condition, with a smooth transition following a hyperbolic
tangent function, to test the 
dependence of our results on the assumption of an initially sharp discontinuity.

This paper also represents one step in our  exploration of
supersonic ambipolar diffusion. The original ZEUS ambipolar diffusion code
of Mac Low et al (1995) has been extended to cover more general
molecular cloud conditions, by including ion mass conservation as opposed to the
ealier assumption of
a fixed ion number density. Hence it is first necessary to test the
behavior of the extended code against standard C-shock solutions (\S 2). The fluid remains as before
isothermal. We thus concentrate on the dynamical aspects. The `cool C-shock
approximation' (Smith \& Brand 1990a) allows us to extract
quantitative predictions for temperature and line emission.

%%%%%%%%%%%%%%%%%%%%%%%%%%%%%%%%%%%%%%%%%%%%%%%%%%%%%%%%%%%%%%%%%%%%%%%%%%%%
%%%%%%%%%%%%%%%%%%%%%%%%%%%%%%%%%%%%%%%%%%%%%%%%%%%%%%%%%%%%%%%%%%%%%%%%%%%%
\section{Framework}
%%%%%%%%%%%%%%%%%%%%%%%%%%%%%%%%%%%%%%%%%%%%%%%%%%%%%%%%%%%%%%%%%%%%%%%%%%%%
%%%%%%%%%%%%%%%%%%%%%%%%%%%%%%%%%%%%%%%%%%%%%%%%%%%%%%%%%%%%%%%%%%%%%%%%%%%%%  
\subsection{Numerical Methods}

For our numerical computations, we use a modified version of the ZEUS
code\footnote{Available for community use by registration with the
Laboratory for Computational Astrophysics at lca@ncsa.uiuc.edu} (Stone
\& Norman 1992a,b).  Ambipolar diffusion was added to ZEUS by Mac Low
et al.\ (1995), who described the basic interface with the ZEUS code.
Summarizing, that work made four approximations: isothermality of
ions, electrons and neutrals, no electron-ion drift, ion density
dependent in power-law fashion on neutral density, and no ion inertia
or pressure.  This allowed us to neglect, respectively, the energy
equation, Ohmic diffusion, and the equations of ion mass and momentum
conservation.  This approach has proved adequate for modelling
protostellar disks in the absence of strong shocks. However, the flow
of ions can be important in C-shocks, for example in the
Wardle instability, which is driven by the flow of ions along buckling
field lines in the shock front. 

\begin{figure} 
\begin{center}
\leavevmode
\psfig{file=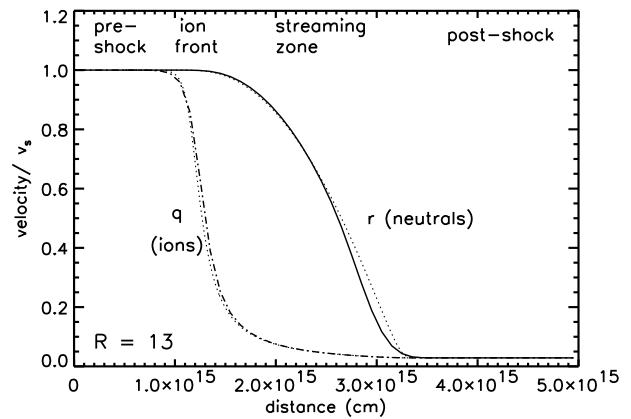,width=8.5cm} 
\caption{The standard transverse-field C-shock reaches this steady state
after 5 10$^{10}$\,s. The grid resolution R$=\Delta x / L_n$, where
$L_n$ is the neutral deceleration length, roughly corresponding to the
shock thickness. Here we show a low-resolution run with R=13 to
emphasize the accuracy of the code. The dotted lines
are the corresponding analytical solutions. The total compression S = 34.86.}
\label{p1_stead}
\end{center}
\end{figure}

Neglect of ion inertia and pressure had allowed the replacement of the
ion momentum conservation equation by an algebraic equation expressing
the balance between Lorentz forces and ion-neutral drag in determining
the drift velocity between ions and neutrals.  This approach is
physically accurate and allows time steps determined by the equivalent
of the Courant condition for ambipolar diffusion, $ \Delta t \le \pi
\gamma \rho_i \rho_n (\Delta x)^2 / |\vecb|^2 $ (Mac Low et
al. 1995). Both T\'oth (1994) and we have found that this
approach can be numerically unstable in the presence of steep velocity
gradients as occur in C-shocks.  However, in the one-dimensional models shown
in this paper, this only occurs for switch shocks, and even there is
not deadly to the computation, so we defer addressing this issue to
our report on multi-dimensional models (Mac Low \& Smith 1997b).

We treat the ions as a separate fluid in the code, using the standard
ZEUS algorithms to update them.  The equations we solve in the current
version of the code are then the neutral and ion continuity equations,
the neutral momentum equation, and the induction equation, as well as
satisfying the zero-divergence criterion:
\begin{eqnarray}
\partial \rho_n/\partial t &=& - \nabla \cdot (\rho_n \vecv_n) \\
\partial \rho_i / \partial t &=& - \nabla \cdot (\rho_i \vecv_i) \\
\rho_n  (\partial \vecv_n / \partial t) &=& - \rho_n (\vecv_n \cdot \nabla)
   \vecv_n - \nabla P_n \nonumber \\
 & & +\gamma \rho_i \rho_n (\vecv_i - \vecv_n) 
   \label{mntmn} \\
(\partial \vecb / \partial t)&=&\nabla \times (\vecv_i \times \vecb) 
    \label{induct} \\
\nabla \cdot \vecb&=&0
\end{eqnarray}
where the subscripts $i$ and $n$ refer to the ions and neutrals, $\rho$,
$\vecv$, and $P$ are density, velocity and pressure for each fluid, 
$\vecb$ is the magnetic field, and $\gamma$ is the collisional 
coupling constant between the ions and neutrals. 

We used the analytic jump conditions for a magnetized shock (e.g.\
Priest 1982) to set the
values of the flow variables on both boundaries needed to hold
the shock stationary on our grid.  These are traditionally called
`inflow' boundary conditions, although on the downstream edge of our
grid the flow is out rather than in.

%%%%%%%%%%%%%%%%%%%%%%%%%%%%%%%%%%%%%%%%%%%%%%%%%%%%%%%%%%%%%%%%%%%%%%%%%%%%
\subsection{Analytical background and parameter definitions}
%%%%%%%%%%%%%%%%%%%%%%%%%%%%%%%%%%%%%%%%%%%%%%%%%%%%%%%%%%%%%%%%%%%%%%%%%%%%
We begin by setting up a one-dimensional J-shock. We generated this solution
from the standard shock jump conditions with magnetic field (e.g. Priest 1982). This is
equivalent to assuming that the ion
fraction was initially high, forcing the ions and neutrals to move at
identical velocity. Given the shock speed,
Alfv\'en speed, temperature and field direction, we can define the 
downstream values in terms of the upstream values.
The supersonic inflow and subsonic outflow boundary conditions are then set so that the
shock front remains stationary on the computational grid. Note that the shock front
thus begins as a truly discontinuous jump rather than being spread out by
numerical viscosity. We also use an initial condition with a smooth transition
proportional to tanh\,(x/L) for a given L to check the importance of the initial
discontinuity.

%%%%%%%%%%%%%%%%%%%%%%%%%%%%%%% TABLE  %%%%%%%%%%%%%%%%%%%%%%%%%%%%%%%%%%%%%%
\begin{table}[htb]
\begin{center}
\caption{Initial and boundary conditions: the standard set.}
\label{initial}
\begin{tabular}{lllll}
\hline
 \\

neutral x-velocity   &  $v_n = rv_s$ &  50 km\,s$^{-1}$ & 1.434 km\,s$^{-1}$  \\
ion x-velocity   &  $v_i = qv_s$ &  50 km\,s$^{-1}$ & 1.434 km\,s$^{-1}$  \\
neutral y-velocity & $v_{ny} = r_yv_s$ &  0  &  0 \\
ion y-velocity & $v_{iy} = q_yv_s$ &  0  &  0 \\
density$^{a}$    &  $n_H$ &  1 10$^5$ cm$^{-3}$& 34.9 10$^5$ cm$^{-3}$ \\

mag. field$^{b}$ &  $B_y$   &  0.3430 mG  & 11.96 mG  \\

ion abundance$^{c}$ & $\chi$ &  $10^{-6}$ &  $10^{-6}$ \\ \\
\hline
\end{tabular}
\end{center}
$^{a}${This is the total hydrogen nuclei number density.}\\
$^{b}${The initial upstream Alfv\'en speed is 2 km\,s$^{-1}$.}\\
$^{c}${The number of ions relative to $n_H$.}
\end{table}

%%%%%%%%%%%%%%%%%%%%%%%%%%%%%%%%%%%%%%%%%%%%%%%%%%%%%%%%%%%%%%%%%%%%%%%%%%%%%%%

 We start with a standard set of initial conditions. This standard set
represents a high Alfv\'en number  shock ($M_a = 25$) with a
transverse magnetic field. In Table 1 we
present all the physical conditions of the models shown in Figures\,1-4, apart from
the  sound speed of 100 cm s$^{-1}$, helium abundance of 0.1, ion
mass $m_i = 10m_p$, neutral mass $m_n = 7m_p/3$ (assuming a fully
molecular gas), and the coupling constant $\gamma$ = $<{\sigma}w>/(m_i+m_n)$ = 9.21
10$^{13}$ cm$^3$\,s$^{-1}$\,gm$^{-1}$, corresponding to a fixed momentum transfer rate coefficient of 
$<{\sigma}w> = 1.9\,10^{-9}$\,cm$^3$\,s$^{-1}$ (a rather uncertain parameter).

We then determine the corresponding steady-state C-shock solution. We 
describe this with the variables $r(x,t) = v_n/v_s$, $q(x,t) = v_i/v_s$ and the fixed temperature $\tau = kT_n/(m_nv_s^2)$.
Hence $r = q = 1$ upstream and the neutral and ion compression ratios
are $1/r$ and $1/q$, respectively. Assuming
ion conservation and isothermality, the steady-state C-shock solution is
then fully described by the total momentum equation
\begin{equation}
 r + \frac{\tau}{r} + \frac{1}{2M_a^2q^2} = 1 + \frac{1}{2M_a^2}
\label{moment}
\end{equation}
 and the drag on the ions
\begin{equation}
L_n\frac{dq}{dx} = \frac{q^2(q-r)}{r}M_a 
\label{moment1}
\end{equation}
(Smith \& Brand 1990a) where $L_n = v_a/(\gamma\rho_i)$  is the neutral deceleration scale. For the `neutralised' isothermal C-shock, in which the
ion abundance is a constant, the momentum conservation condition is
unaltered and the drag relation is then
\begin{equation}
L_n\frac{dq}{dx} = \frac{q^3(q-r)}{r}M_a. 
\label{moment2}
\end{equation}
Full steady-state solutions are derived by numerically solving these simple differential equations to any desired accuracy.

%%%%%%%%%%%%%%%%%%%%%%%%%%%%%%%%%%%%%%%%%%%%%%%%%%%%%%%%%%%%%%%%%%%%%%%%%%%%
\subsection{Scaling}
%%%%%%%%%%%%%%%%%%%%%%%%%%%%%%%%%%%%%%%%%%%%%%%%%%%%%%%%%%%%%%%%%%%%%%%%%%%%

Besides the length scale $L_n$, we define the ion
deceleration scale $L_i = L_n/M_a$. A further scale defines the
ion-neutral interaction region. In cool shocks, the maximum streaming speed
occurs for $q = 1/M_a^{2/3}$ (on putting t = 0 in Equation \ref{moment}). The value of the maximum 
is $(r-q)_{max} = 1 + M_a^{-2} - 3M_a^{-2/3}/2$. It follows that high streaming speeds are maintained
over a length scale $L_{str} \sim L_n/M_a^{1/3}$. Finally note that the
length scale defined by Wardle (1990, 1991a,b) is $L_{shk} =
\sqrt{2}\,L_n$.
We define here what proves to be a better measure of the shock length for transverse shocks: $L_{sm} = L_n/(r-q)_{max}$. This enables low $M_a$ shocks to be also included

The time-dependent shock flow pattern is fully determined by four control
parameters: the Alfv\'en number $M_a$, the Mach number $M$, the ion
fraction and the initial field orientation. These are the only parameters
which remain after scaling
equations (1--5) to the length scale $L_n$ and the time scale
$t_{flow}  = L_n/v_a$. 
Therefore only one parameter is necessary to describe the whole class of transverse, low-ionisation, cold flows: a single flow simulation with a fixed  Alfv\'en number is relevant to a wide range of conditions.  

One must, however, ensure that the thermal pressure gradients
remain small. This is  easily achievable when close to  the steady state
since then the pressure gradient term $\tau/r = c_s^2/(v_nv_s)$ is limited by the maximum compression ($S \sim \sqrt{2}M_a$ for high $M_a$) to the value ($c_s^2/v_a^2)/S$,
where $c_s$ is the sound speed.
This is expected to be well below unity (and therefore ignorable) in
strong shocks in molecular clouds. However, in time-dependent flows in general, depending on the
imposed conditions,  neutral speeds may approach zero and even an isothermal flow
may possess high pressure gradients. The present simulations then require
holding both the Mach and Alfv\'en numbers fixed.

Furthermore, the freedom to scale parameters while holding $M_a$ fixed
becomes 
physically invalid if the ion fraction depends
on the other shock parameters. This limits the scaling regime to flows
with ion-neutral streaming speeds less than $\sim$ 42 km\,s$^{-1}$ to
avoid runaway ionisation (Draine et al 1983, Smith \& Brand 1990a).

The flow patterns presented below can thus be considered quite general
provided the above rules are not violated.  No scaling was attempted,
however, when calculating the emission line properties, in order to
avoid possible mis-matching when piecing together the results from
separate models for the emission from the J-shock and the continuous
section.

%%%%%%%%%%%%%%%%%%%%%%%%%%%%%%%%%%%%%%%%%%%%%%%%%%%%%%%%%%%%%%%%%%%%%%%%%%%%
\subsection{Test of accuracy}
%%%%%%%%%%%%%%%%%%%%%%%%%%%%%%%%%%%%%%%%%%%%%%%%%%%%%%%%%%%%%%%%%%%%%%%%%%%%

In our time-dependent computations, J-shocks  indeed evolve to the 
analytical, steady C-shock solution to within our numerical accuracy. 
Figure~\ref{p1_stead}
presents the flow parameters at a time $5\,10^{10}$\,s. 
The length scale is $L_n$ = 1.3\,10$^{15}$\,cm and the flow time scale is
$t_{flow}$ = $L_n/v_a$ = $6.5\,10^9$\,s. We find that it takes several flow time
scales before the flow pattern finally settles down to the 
analytical steady state solution.

\begin{figure} 
\begin{center}
\leavevmode
\psfig{file=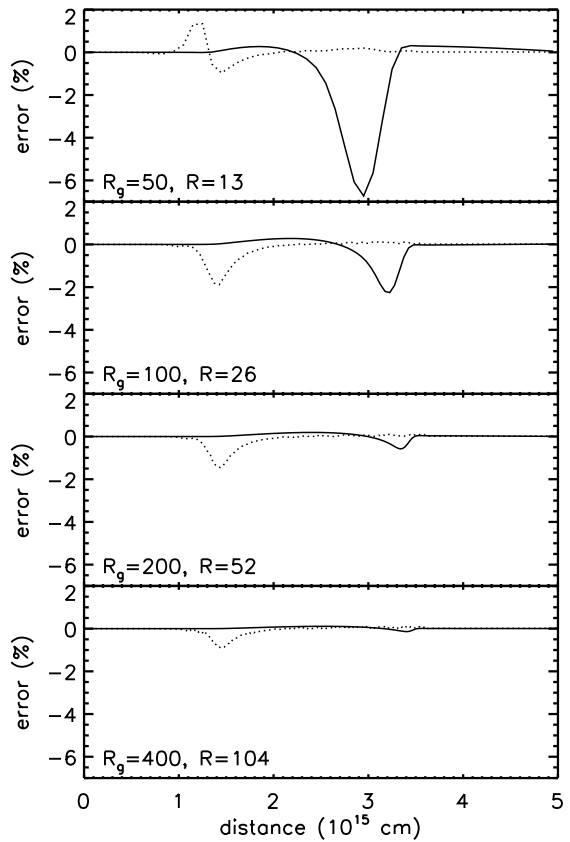,width=9cm} 
\caption[junk]{The errors in the final configurations for simulations
of the standard C-shock at the four indicated resolutions $R = \Delta
x /L_n$.  The number of zones in the full grid is given by $R_g$. The
solid line is the momentum error and the dotted line is the streaming
error.}
\label{p1_error}
\end{center}
\end{figure}

Figure~\ref{p1_error} demonstrates the dependence of the
time-dependent solutions on the grid resolution. Even a grid size of R$_g$ = 50,  with the number of zones across the shock 
 R = ${\Delta}x/L_n$ = 13, is sufficient for dynamical purposes with the major error occurring in the final deceleration zone (at about 3\,10$^{15}$ cm).
The error functions analysed in Figure~\ref{p1_error} are the momentum error (solid
lines) given by 
\begin{equation}
\epsilon_m(x) = \frac{r + 1/(2M_a^2q^2)}{1 + 1/(2M_a^2)}
\label{error1}
\end{equation}
and the streaming error (dotted lines) given by
\begin{equation}
\epsilon_s(x) = \left[r_s-q_s\right] - r + q
\label{error2}
\end{equation}
where  $\left[r_s-q_s\right]$ is the analytical steady streaming solution. The expected second order convergence  (Mac Low et al 1995) occurs.

%%%%%%%%%%%%%%%%%%%%%%%%%%%%%%%%%%%%%%%%%%%%%%%%%%%%%%%%%%%%%%%%%%%%%%%%%%%%
%%%%%%%%%%%%%%%%%%%%%%%%%%%%%%%%%%%%%%%%%%%%%%%%%%%%%%%%%%%%%%%%%%%%%%%%%%%%
\section{The Evolution from J-shock to C-shock}
%%%%%%%%%%%%%%%%%%%%%%%%%%%%%%%%%%%%%%%%%%%%%%%%%%%%%%%%%%%%%%%%%%%%%%%%%%%%
%%%%%%%%%%%%%%%%%%%%%%%%%%%%%%%%%%%%%%%%%%%%%%%%%%%%%%%%%%%%%%%%%%%%%%%%%%%%

\subsection{The four stages for transverse shocks}

We identify four  stages in the evolution from a J-shock to a C-shock. Each stage lasts approximately 10 times longer than the previous one.

\begin{figure} 
\begin{center}

\leavevmode
\psfig{file=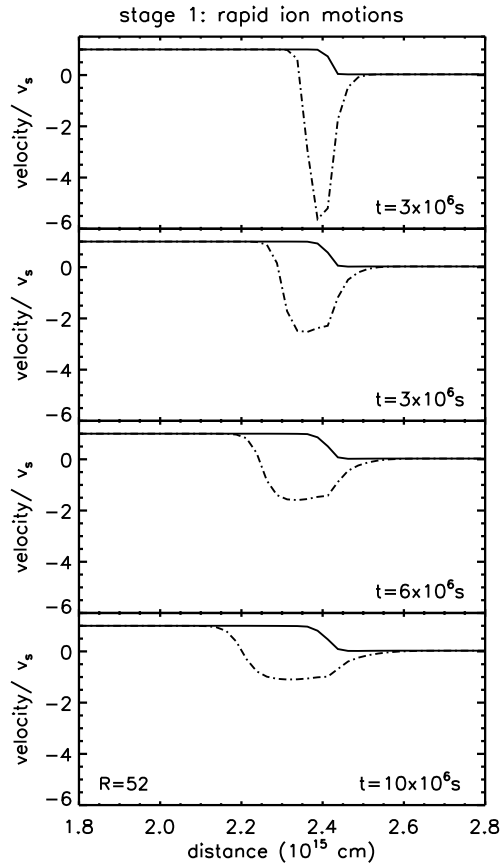,width=9cm} 
\caption[junk]{The initial phase in the J to C-shock transition
involves rapid ion motions.  The neutral (solid line) and ion
(dash-dotted line) velocities are shown.  Note the distance scale has been expanded,
so that the neutral jump shock, spread over 2--3 zones by numerical
viscosity, appears rather broad.}
\label{p1_stage1}
\end{center}
\end{figure}
\begin{figure} 
\begin{center}

%\leavevmode
\psfig{file=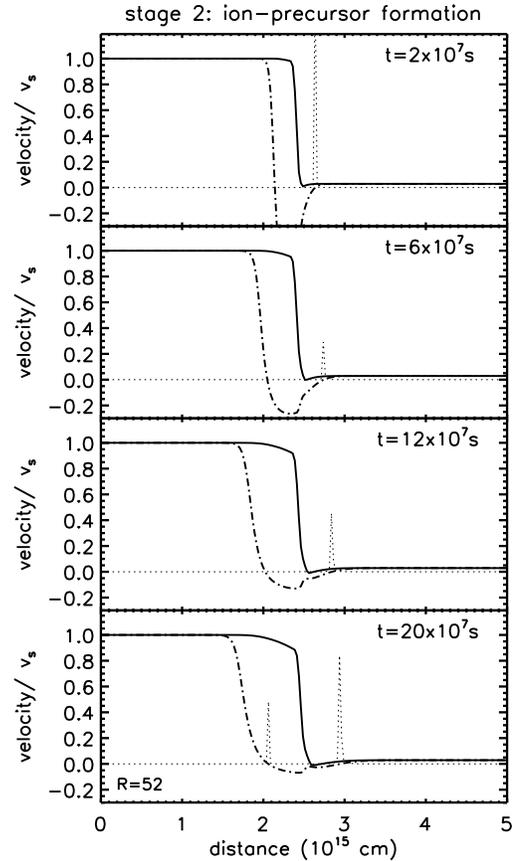,width=9cm} 
\caption[junk]{The second phase in the J to C-shock transition. The
ion and neutral fronts separate at speeds in the range 10--100
km$^{-1}$.  The neutral (solid line) and ion (dash-dotted line)
velocities are shown, as well as $0.001 \epsilon_m$ (dotted line),
where the momentum error $\epsilon_m$, given by equation~\ref{error1},
tracks the deviation from the steady-state solution. }
\label{p1_stage2}

\end{center}
\end{figure}
\begin{figure} 
\begin{center}

%\leavevmode
\psfig{file=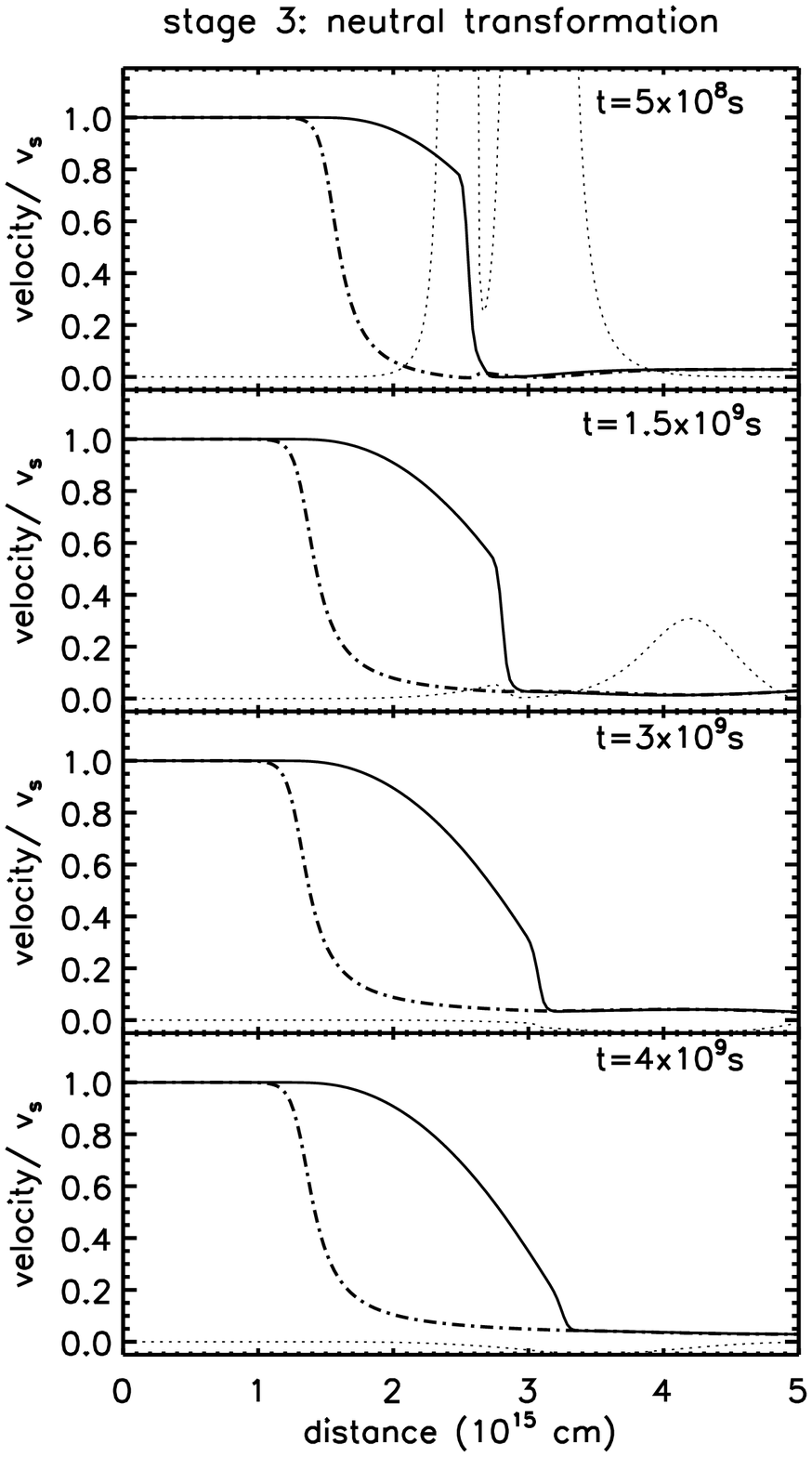,width=9cm} 
\caption[junk]{The neutral jump shock finally disappears and the shock
width further increases in Stage 3. The neutral (solid line) and ion
(dash-dotted line) velocities are shown, as well as $0.1 \epsilon_m$
(dotted line), where the momentum error $\epsilon_m$ is given by
equation~\ref{error1}.  }
\label{p1_stage3}
\end{center}
\end{figure}

\begin{itemize}

\item Stage 1 involves rapid ion motions. It is characterised
by high-speed, ion-magnetosonic wave motions over a short time. During
the first $0.002 t_{flow} \sim 10^7$\,s, the
precursor moves upstream at a speed of order 1000 km\,s$^{-1}$, the ion-magnetosonic  speed. As shown in Figure~\ref{p1_stage1}, the ions
move upstream (negative velocities), away from the original shock front, with
a speed well above the shock speed. The ion compression wave is separated from the neutral front by an ion expansion wave, as expected in this shock-tube
experiment. 

\item During Stage 2 the ion front and the neutral front separate over a
time of $0.03 t_{flow} \sim 2\,10^8$\,s, at intermediate speeds (Figure~\ref{p1_stage2}). The ion precursor develops fully. The neutrals possess a weak
precursor but are still hardly altered. 

\item  Stage 3 represents the approach to a C-type structure, as the
neutrals evolve towards a fully continuous flow. The changes now occur
at the Alfv\'en speed of 2\,km\,s$^{-1}$ over a time of $0.6 t_{flow}
\sim 4\,10^9$\,s.

\item During  Stage 4  only small changes to the flow pattern occur. It takes several Alfv\'en speed crossing times before the steady state solution is
finally reached, as described in the previous section (see Fig.~\ref{p1_stead}). The final length scale is $\sim 1.6\,10^{15}$cm,
approximately equal to $L_{sm}$.

\end{itemize}

%%%%%%%%%%%%%%%%%%%%%%%%%%%%%%%%%%%%%%%%%%%%%%%%%%%%%%%%%%%%%%%%%%%%%%%%%%%%
\subsection{Low Alfv\'en number shocks}
%%%%%%%%%%%%%%%%%%%%%%%%%%%%%%%%%%%%%%%%%%%%%%%%%%%%%%%%%%%%%%%%%%%%%%%%%%%%

Strong magnetic fields cushion a shock by broadening the transition region and reducing the drag heating in the shock front. This is especially important for
C-shocks since the cushioning changes the character of the whole transition by reducing the ion-neutral streaming speed whereas for J-shocks it is usually only felt in the downstream compressed gas. Hence we call these C-shocks `shock absorbers' (Smith et al 1991). This  cushioning has
observational implications: molecules are not so easily
destroyed by ion collisions because the ion-neutral streaming speed is a
relatively small fraction of the shock speed. In order to model a
shock absorber, we 
simply reduce the shock speed from the standard case by a factor of 10
to simulate a $M_a = 2.5$ shock.

\begin{figure} 
\begin{center}
\leavevmode
\psfig{file=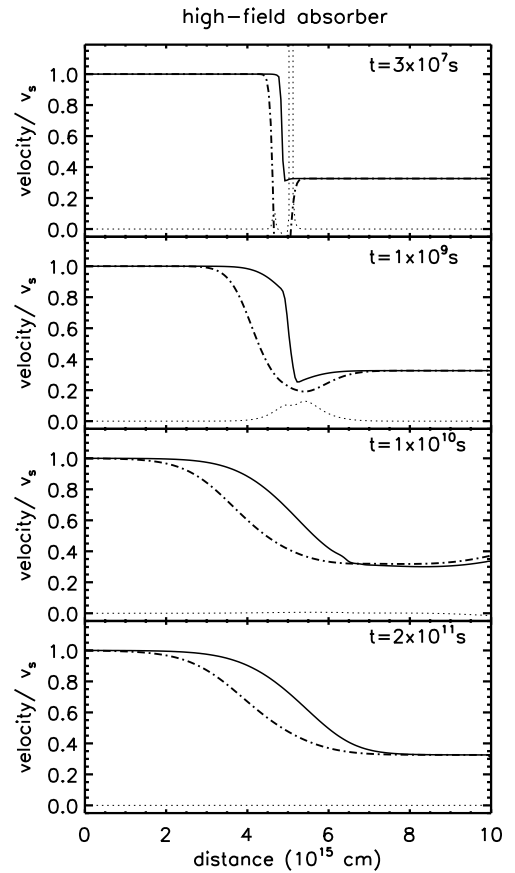,width=9cm} 
\caption[junk]{At low Alfv\'en numbers, the ions and neutrals gently
stream against each other. In this model, the shock speed $v_s = $
5\,km\,s$^{-1}$, the neutral Alfv\'en number $M_a = 2.5$, the Mach
number $M = 50$, and the compression $S = 3.07$. Note the distance
scale has been doubled. The grid size $R_g = 200$, so the resolution R
= 110 zones in $L_n$. The neutral (solid line) and ion (dash-dotted
line) velocities are shown, as well as $0.1 \epsilon_m$ (dotted line),
where the momentum error $\epsilon_m$ is given by
equation~\ref{error1}. }
\label{p1_absorber}
\end{center}
\end{figure}

We find that all the transition stages are identifiable but last several times 
longer (Figure~\ref{p1_absorber}). Indeed it takes $\sim 10^4$ years to reach the steady state, comparable to the age of many outflows. Note that the total 
compression is only about 3 in the example shown. The final total length of the
shock is $\sim 4\,10^{15}$\,cm. It is clear that  $L_{sm}$ is indeed an accurate measure of the shock length scale.

%%%%%%%%%%%%%%%%%%%%%%%%%%%%%%%%%%%%%%%%%%%%%%%%%%%%%%%%%%%%%%%%%%%%%%%%%%%%
\subsection{Oblique field}
%%%%%%%%%%%%%%%%%%%%%%%%%%%%%%%%%%%%%%%%%%%%%%%%%%%%%%%%%%%%%%%%%%%%%%%%%%%%

\begin{figure} 
\begin{center}
%\leavevmode
\psfig{file=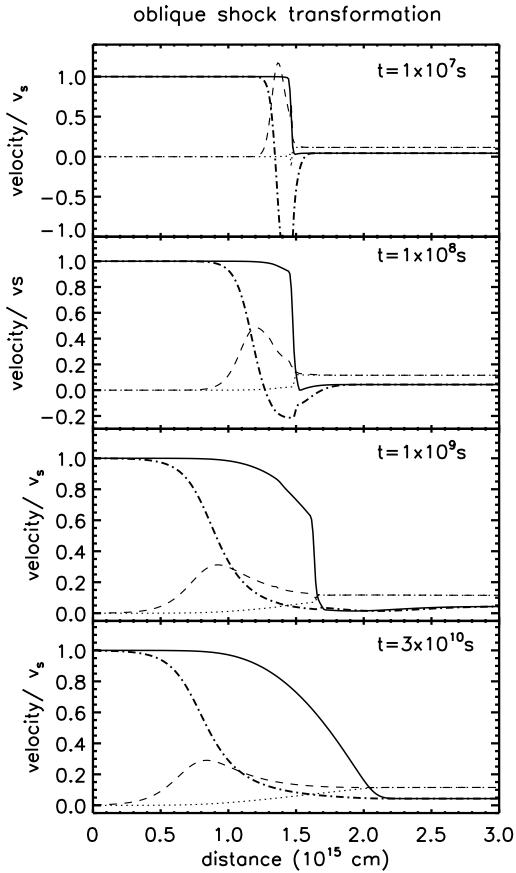,width=9cm} 
\caption[junk]{An oblique shock with the upstream magnetic field at an
angle of 30$^\circ$ to the shock velocity.  The ion (dash-dotted line)
and neutral (solid line) velocity components along the flow
direction, qv$_s$ and rv$_s$, are shown as before, as well as the
transverse speeds 
q$_y$v$_s$ (dashed line) and r$_y$v$_s$ (dotted line). The upstream transverse
speeds were set to zero. Here R$_g$ = 200, and R = 109. The shock
speed is 20\,km\,s$^{-1}$, Mach number $M = 200$ and $M_a = 10$, giving a total compression S = 23.1.}
\label{p1_obl}
\end{center}
\end{figure}

Oblique shocks develop from the initial state to a steady-state,
oblique C-shock without any surprises. By oblique shock, we specifically
mean that the magnetic field is inclined to the shock velocity, but
not so far inclined 
that intermediate shock solutions become possible (Smith 1993c; see below). In this regime, the ion and neutral speeds along the flow
direction, qv$_s$ and rv$_s$, behave as before, as shown in
Fig.~\ref{p1_obl}. The ion transverse speed  q$_y$v$_s$  exceeds the
neutral transverse speed r$_y$v$_s$. Note also that the ion transverse
speed passes through a maximum, and
that the shock is narrower than the transverse solution. Thus the grid
size was reduced to   preserve approximately the same numerical resolution.

%%%%%%%%%%%%%%%%%%%%%%%%%%%%%%%%%%%%%%%%%%%%%%%%%%%%%%%%%%%%%%%%%%%%%%%%%%%%
\subsection{Near-parallel field}
%%%%%%%%%%%%%%%%%%%%%%%%%%%%%%%%%%%%%%%%%%%%%%%%%%%%%%%%%%%%%%%%%%%%%%%%%%%%

When the magnetic field is quasi-parallel, the flow does not reduce to
the hydrodynamical equivalent, as is often assumed. Even as the
transverse field approaches zero, the flow should approach a
switch shock solution. This is due to the strong cooling in
molecular shocks, or, here, the assumption of isothermality (Smith 1993a,c). 
Although it is
possible that in complex, multi-dimensional, shock configurations a mixture of the
three possible shock solutions  can be maintained at field angles  less than $\sim
1/M_a$ radians,  the switch-type solution
is the only evolutionary solution in plane-parallel flows (Kennel 1988).
\begin{figure} 
\begin{center}
%\leavevmode
\psfig{file=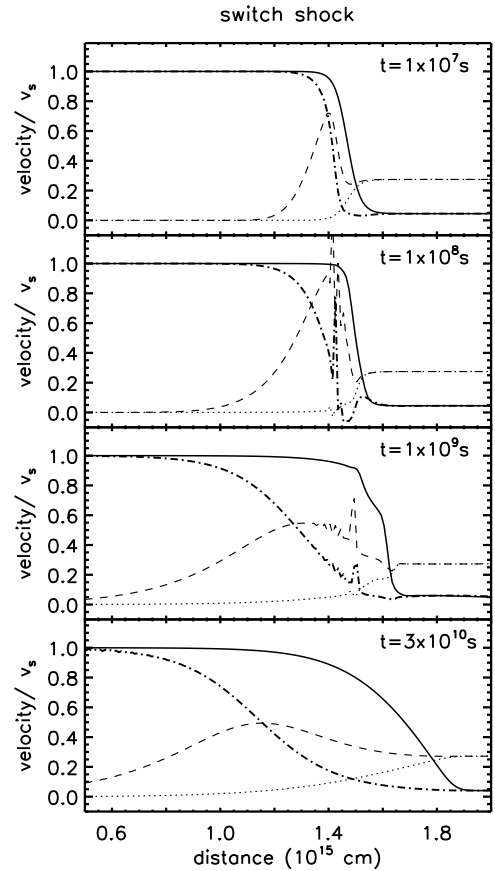,width=9cm} 
\caption[junk]{The evolution of a switch shock structure with shock
velocity $v_s = 10$ \,km\,s$^{-1}$, magnetic
field at an angle of 2$^\circ$ to $v_s$, a compression of
$S = 22.31$, and  $M_a = 5$.  The parallel field is
constant (3.43~10$^{-4}$\,G) while the transverse field is
`switched on' from 1.20~10$^{-6}$\,G to 2.37~10$^{-3}$\,G. The ion and
neutral velocity components are shown as described in
Fig.~\ref{p1_obl}. The grid size is R$_g$ = 300 zones, giving a
resolution R = 121. }
\label{p1_switch2}
\end{center}
\end{figure}

To compute switch shocks we had to solve several problems. 
First, we had to set up an appropriate grid.  If we
take a magnetic field
angle of 2$^\circ$ to the flow direction, our standard parameters
lead to resolution problems: the neutral shock width is reduced to
$\sim L_n/M_a$, but, in contrast, the ion precursor extends far
upstream.  We solve this problem simply by extending the grid upstream
using a ratioed grid, and increasing the total number of zones.  This
problem
can still be seen in Fig.~\ref{p1_switch2}, where we have reduced $M_a$ to 5
($v_s$ = 10\,km\,s$^{-1}$). The whole ion precursor, as evident
in the upstream transverse ion speed, is still not shown. This
precursor will not significantly warm the gas, however, as the heating
occurs where the drag is strong and the neutrals decelerate (see Fig.~4. of
Smith (1993b)). 

A second problem
is inherent to the ion treatment in this code: steep
ion velocity gradients generate numerical instabilities (Mac Low \& Smith 1997b). Unfortunately, such steep gradients are  exactly what occurs in flows set up along the
field lines. Hence, at early times, up to $\sim 6\,10^8$\,s, 
numerical instability is evident within the ion flow. The instability
is damped as the 
gradients diminish. This instability
does not occur when, instead of discontinuous initial conditions, we use a hyperbolic tangent 
function with a width of 1\,10$^{14}$\,cm, and we get nearly identical
results. 

The tranverse ion motions are found to have a 
maximum, as was shown to be true for shocks with $M_a$ $> 3{\surd}2/2$, in
the steady cold solutions of Smith (1993a).

We have also set up the quasi-hydrodynamic, or intermediate Type II
boundary and initial jump conditions. In intermediate shocks, the
post-shock sound speed is subsonic. As predicted by analytic theory,
this flow remained J-type and steady. It is clear that a C-type flow
should not occur since the ions, without the inertia of the magnetic
field, are now strongly tied to the neutrals. Such Type II J-shocks in
multiple dimensions could degenerate into a fast (switch type) and
slow shock combination; exactly what is possible depends on the shock
configuration and applied conditions (see Smith 1993c).  We find,
however, that slow shocks (e.g. Smith 1993c) are also restricted to
the J-shock variety.  In this case, the sub-Alfv\'enic motions tie
together the ions and neutrals.

\begin{figure} 
\begin{center}
%\leavevmode
\psfig{file=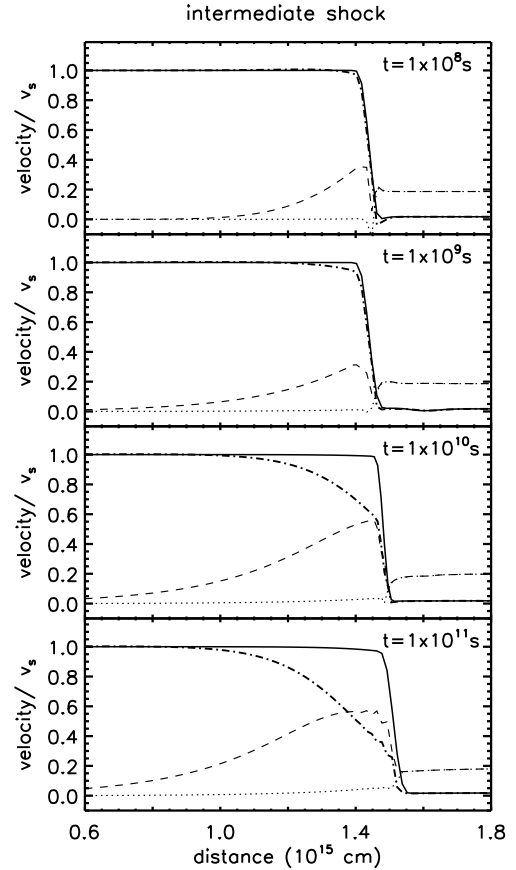,width=9cm} 
\caption[junk]{The evolution of an intermediate shock structure with
the field at an angle of 6$^\circ$ to the shock velocity, a compression of 57.14, Mach
number M = 10, and neutral Alfv\'en velocity M$_a = 5$. The parallel field is 
constant (0.341~mG) while the transverse field flips from 0.0362~mG through zero
to -1.57~mG. The four lines are as described in Fig.~\ref{p1_obl}, and
the grid size is R$_g$ = 300 zones.}
\label{inter3}
\end{center}
\end{figure}
Type I intermediate shocks are similar to switch shocks except the small
transverse field
reverses direction within the shock layer.  We find the evolutionary
behaviour follows that of the switch shock, but at a slower pace
(approximately 3 times slower for the standard conditions with the field
initially at 1$^\circ$ to the shock normal.) 

We conclude that a continuous
range, from C-type to J-type occurs, when the boundary conditions are altered
from switch, via Type I and Type II intermediate, to hydrodynamic type. This
is emphasized through Fig.~\ref{inter3}, which is an example of an
intermediate shock close to the I-II transition border. Here the jump
shock remains while a partial ion/magnetic precursor develops. Note that here also the transverse ion Alf\'en waves propagate far upstream (the complete
computational grid is not shown).

%%%%%%%%%%%%%%%%%%%%%%%%%%%%%%%%%%%%%%%%%%%%%%%%%%%%%%%%%%%%%%%%%%%%%%%%%%%%
\subsection{Neutralisation in transverse shocks}
%%%%%%%%%%%%%%%%%%%%%%%%%%%%%%%%%%%%%%%%%%%%%%%%%%%%%%%%%%%%%%%%%%%%%%%%%%%%

A class of physical conditions produce another type of
shock structure in which the ions are not conserved. A
particularly simple  case, in which the ion fraction remains fixed
throughout, is considered here.  This can occur when the recombination
time is short, such as for molecular ions, with rate coefficients $R_n
\sim 10^{-7}~$cm$^3~$s$^{-1}$ rather than atomic ions; and when the
shocks are wide enough for
neutralisation reactions to occur (Flower et al 1996). Since the shock width is
inversely proportional to the ion density, neutralisation is
determined solely by the recombination
rate coefficient and Alfv\'en number.  The ratio of shock dynamical to
recombination times is $\sim 100(R_n/10^{-7}~$cm$^3~$s$^{-1})/M_a$. 

\begin{figure} 
\begin{center}
%\leavevmode
\psfig{file=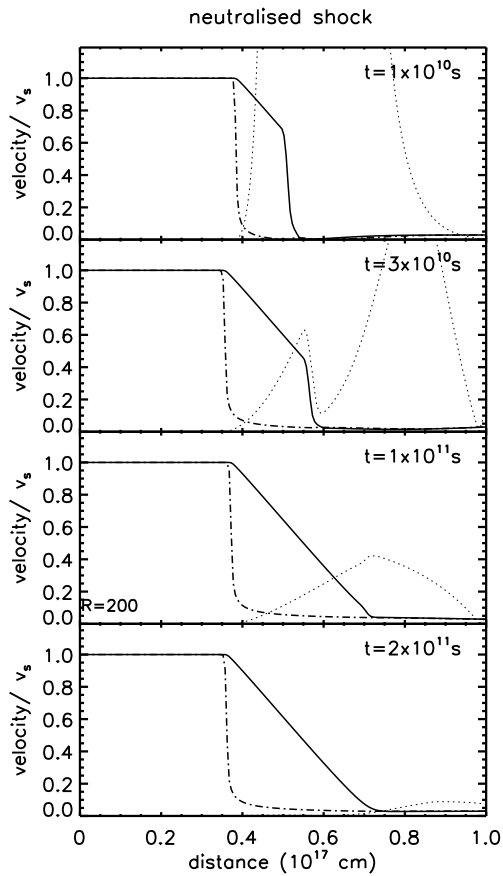,width=9cm} 
\caption[junk]{The flow patterns for a neutralised shock transition. All
parameters are as chosen for the standard transverse shock except the
ion
density is held at its initial value. Note
the broader shock. The grid size is R$_g$ = 200 zones, so the
resolution is only R = 68 zones in the shock width. The neutral (solid line)
and ion (dash-dotted line) 
velocities are shown, as well as the momentum error $ \epsilon_m$ (dotted line).
 }
\label{p1_neutral}
\end{center}
\end{figure}

We find that neutralised shocks evolve from the imposed J-type initial state
directly to C-type. The neutral jump section is preceded by a ramp of
growing amplitude, as shown in Figure~\ref{p1_neutral}. The final
C-shock flow pattern is now the ramp 
structure in the neutrals, preceded by the rapid ion braking.  In this
zone of rapid ion braking and compression the recombinations will be
hard pressed to 
suppress a rise in ion density.  This may lead to a narrow peak in the
ion density in reality. 

The time to reach  steady state in a neutralised shock is
much longer than for the other shock types. Almost 10$^4$ years is required for
the conditions chosen, comparable with  the duration of the Class 0
stage of a protostar. Shock widths are correspondingly
larger. Furthermore, the age and shock width are 
inversely proportional to the ion density, as we have checked through
further simulations.
Hence, low-ionisation  shocks may never actually reach a steady state.

%%%%%%%%%%%%%%%%%%%%%%%%%%%%%%%%%%%%%%%%%%%%%%%%%%%%%%%%%%%%%%%%%%%%%%%%%%%%
%%%%%%%%%%%%%%%%%%%%%%%%%%%%%%%%%%%%%%%%%%%%%%%%%%%%%%%%%%%%%%%%%%%%%%%%%%%%

\section{Observables}
%%%%%%%%%%%%%%%%%%%%%%%%%%%%%%%%%%%%%%%%%%%%%%%%%%%%%%%%%%%%%%%%%%%%%%%%%%%%
%%%%%%%%%%%%%%%%%%%%%%%%%%%%%%%%%%%%%%%%%%%%%%%%%%%%%%%%%%%%%%%%%%%%%%%%%%%%

Shock waves in a magnetized molecular gas will be continuously
changing, both through 
the C-shock instability (Wardle 1990, Neufeld \& Stone 1997, Mac Low
\& Smith 1997a,b) and through interaction with 
spatial perturbations in density, ion fraction and magnetic field.  So
how can we recognise the evolving shocks described in this paper? To
calculate the emission properties as a function of time, 
we must first estimate the temperature distribution through the shock.
In the continuous sections this can be done using the `cool C-shock'
approximation (Smith \& Brand 1990a), provided 
that the temperature remains low so that the thermodynamics and
magnetohydrodynamics are decoupled.  Quantitatively, we require that
the temperature T $\ll$ m(H$_2$)$v_s^2$/k. The 
low isothermal temperature adopted ensures this condition in our numerical
computations, and strong H$_2$ cooling often ensures it in real
molecular clouds.

The jump sections are dealt with separately: the jump parameters are transferred
into the J-shock code of Smith (1994a), the column densities of ions
in each excited level are calculated, and then
added to the column densities derived from the continuous section in
the manner described below.

The temperature profile in the continuous section is determined by the
local balance of ion-neutral
frictional heating with molecular cooling (see Smith 1993b). Simplified cooling functions are adopted consistent with the limitations of the
calculations already introduced (in particular the simplified drag formula). Here we restrict the illustrative results to that of H$_2$ cooling, with the H$_2$ molecule in local thermodynamic equilibrium. The cooling function
is extracted from Smith (1993b): 
\begin{equation}
\Lambda(T) =
(4.2\,10^{-31}\mbox{~erg~s}^{-1}\mbox{~cm}^{-3}) n({\rm H}_2)T^{3.3}.
\end{equation}
It is
straightforward to show that the temperature is then given by T =
2.89n$_i$v$_{in}^2$ where v$_{in}$ is the ion-neutral streaming
speed. 

Column densities of molecules in excited upper levels can be computed from
\begin{equation}
N_j = g_jN(H_2) Z(T) \exp(-T_j/T),
\end{equation}
assuming that the rotational levels are in LTE at the temperature T,
where the partition function is
\begin{equation}
Z(T) = 0.024\,T\,\left[1 - \exp(-6000/T)\right]^{-1},
\end{equation}
and g$_j$ are the statistical weights.  Although we assumed the
rotational levels are in LTE, we allowed
the vibrational levels to fall out of LTE using the method described
by Suttner et al.\ (1997).
Line strengths $I_j$ are then calculated from the column of gas N$_j$(x)
in the transition's upper energy level: 
\begin{equation}
I_j = \int (hc/\lambda_j)A_j N_j(x)dx,
\end{equation}
where A$_j$ is the radiative 
coefficient and $\lambda_j$ the wavelength of the transition. 
\begin{figure} 
\begin{center}
%\leavevmode
\psfig{file=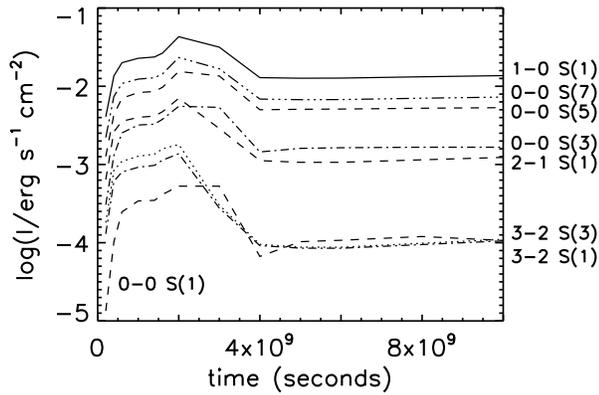,width=10.6cm} 
\caption[junk]{Line intensities versus time for a 30 km\,s$^{-1}$
transverse shock model. The shock is assumed to be observed face on
(thus the columns are calculated along a line of sight parallel to the
shock velocity). The physical data are given in
Table~\ref{linetable}.}
\label{intensity}
\end{center}
\end{figure}

%%%%%%%%%%%%%%%%%%%%%%%%%%%%%%% TABLE  %%%%%%%%%%%%%%%%%%%%%%%%%%%%%%%%%%%%%%
%%%%%%%%%%%%%%%%%%%%%%%%%%%%%%% TABLE  %%%%%%%%%%%%%%%%%%%%%%%%%%%%%%%%%%%%%%
\begin{table}[htb]
\begin{center}
\caption{Physical data for selected molecular hydrogen emission lines,
observable in either
the K-band (2$\mu$m-2.4$\mu$m) or by the SWS of the Infrared Space Observatory
(0-0 transitions). Their intensities are plotted in Fig.~\ref{intensity}.}
 \label{linetable}
 \begin{tabular}[t]{lrrll}\hline\hline

Line & $\lambda_j^{a}$ &  $T_j^{b}$ & 
       $A_j^{c}$     &  $g_j^{d}$ \\
\hline \\
1-0 S(1) &  2.1213 &  6951. &   3.47 &   21.    \\   
3-2 S(3) &  2.2008 &  19086. &   5.63  &  33.   \\   
2-1 S(1) &  2.2471 &  12550. &   4.98 &   21.     \\   
3-2 S(1) &  2.3858 &  17818.  &  5.14  &  21.    \\  
0-0 S(7) &  5.5110  &  7199.   &  2.00   &    57. \\   
0-0 S(5) &  6.9100  &  4587.  &   0.588   &   45. \\   
0-0 S(3) &  9.6650  &  2504.   &  0.0984   &  33. \\   
0-0 S(1) & 17.0350  &  1015. &    0.00476 &   21. \\   \\
\hline
\end{tabular}
\end{center}

$^{a}${Wavelength in $\mu$m}\\
$^{b}${Excitation temperature in K}\\
$^{c}${Einstein A-values for radiative deexcitation in 10$^{-7}$\,s$^{-1}$}\\
$^{d}${Statistical weights with ortho/para ratio of 3}
\end{table}
%%%%%%%%%%%%%%%%%%%%%%%%%%%%%%% TABLE  %%%%%%%%%%%%%%%%%%%%%%%%%%%%%%%%%%%%%%

\begin{figure} 
\begin{center}
%\leavevmode
\psfig{file=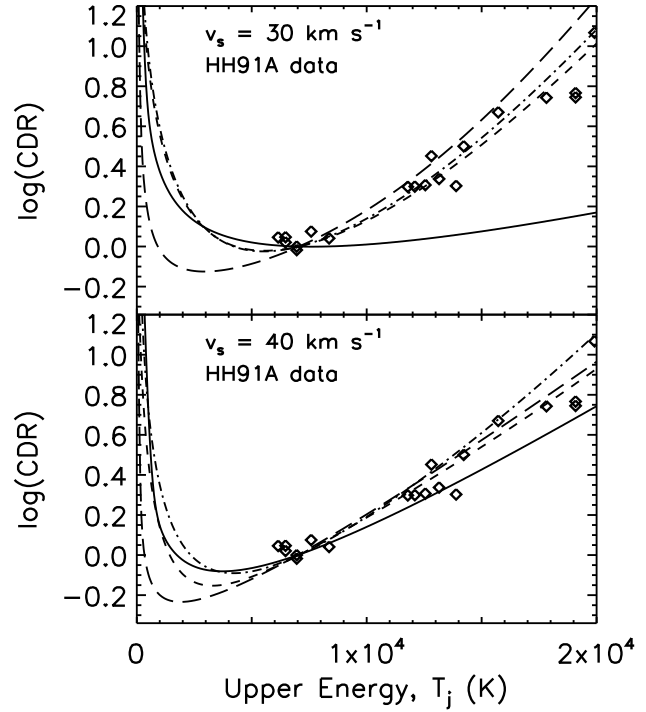,width=8.6cm} 
\caption[junk]{CDR diagrams (see text) for the standard transverse
field model with the 
indicated shock speed and evolution times of 2\,10$^8$\,s (long dash),
10$^9$\,s (dot-dash),  2\,10$^9$\,s (dash),  10$^{10}$\,s (full
lines).   The data points displayed here were calculated by Smith
(1994b) from the intensities observed by Gredel et al (1992). Only
positive line detections are shown here.} 
\label{p1_cdr}
\end{center}
\end{figure}
The time dependence of eight lines observable either in the K-band or
with the SWS of the Infrared Space Observatory are shown in
Fig.~\ref{intensity}, and the physical parameters of these lines are given in
Table~\ref{linetable}. The columns of gas are displayed as column
density ratios (CDRs) in 
Fig.~\ref{p1_cdr}. A CDR for H$_2$ is the column of gas in the energy level
T$_j$ divided by the column of gas in the v = 1, J = 3 level that
generates the 1-0 S(1) line, further normalised by the factor
exp(T$_j$/2000K) to remove the strong temperature dependence. We 
thus are comparing column densities to those of a slab of molecular
gas at a temperature of 2000~K.  These CDR diagrams are an accurate means of displaying
the H$_2$ excitation over a broad range of energy levels (see also Mac Low \& Smith 1997b).
A 30 km\,s$^{-1}$ model was chosen for display purposes since this
produces gas up to a maximum temperature of $\sim 2600\,$K within the 
final steady C-shock. The CDRs for a hotter 40 km\,s$^{-1}$ model are
also shown. The following stages are recognisable.
\begin{itemize}
\item The structure begins as a fast dissociative jump shock with
low luminosity in these infrared lines (Fig.~\ref{intensity}). The
ion-magnetic precursor 
develops rapidly in the early stages, producing a sharp increase in
line strengths.
\item The jump speed drops below $\sim$ 24 km\,s$^{-1}$ at $\sim 4\,10^8\,$s.
Now the molecules survive the jump.
The molecules are strongly heated in the jump shock producing a high-excitation
spectrum, as best illustrated by the high CDRs during this stage in
the high energy levels. Note that HH91A possesses such a high
excitation spectra, as shown in Figure~\ref{p1_cdr}. 
\item The jump weakens and is unable to excite the molecules after
$\sim 3\, 10^9\,$s. The shock excitation and line intensities are now controlled by the ion-neutral drag and rapidly approach  their final values. The
final excitation depends strongly on the shock parameters.
\end{itemize}

We find no significant variations from the above behaviour for the
other types of shocks. Switches, oblique shocks and intermediate
shocks are of course hotter than the 
equivalent transverse shocks. Neutralised shocks possess similar excitation
signatures as transverse shocks.

%%%%%%%%%%%%%%%%%%%%%%%%%%%%%%%%%%%%%%%%%%%%%%%%%%%%%%%%%%%%%%%%%%%%%%%%%%%%
%%%%%%%%%%%%%%%%%%%%%%%%%%%%%%%%%%%%%%%%%%%%%%%%%%%%%%%%%%%%%%%%%%%%%%%%%%%%
\section{Conclusions}
%%%%%%%%%%%%%%%%%%%%%%%%%%%%%%%%%%%%%%%%%%%%%%%%%%%%%%%%%%%%%%%%%%%%%%%%%%%%
%%%%%%%%%%%%%%%%%%%%%%%%%%%%%%%%%%%%%%%%%%%%%%%%%%%%%%%%%%%%%%%%%%%%%%%%%%%%

We have simulated the evolution of jump shocks into continuous
shocks. We set up boundary and initial conditions appropriate to
the C-shock, but separated the two flow regimes by a discontinuous
diaphragm as in classical shock tube experiments. The questions we
hope to answer, however, are more general: how do shocks behave in a
non-uniform medium?  Do steady-state C-shocks form?  We have indeed
found that, in all cases, analytic steady-state solutions are
approached, given sufficient time.  That time can in the most extreme
cases, however, be comparable to the lifetimes of embedded protostars.

Four stages were identified for most shock types. First, an ion
expansion wave rapidly 
advances into the upstream region. Next, the ion-magnetic precursor forms and
moves gradually upstream, with a
high streaming speed capable of heating the molecules. The neutrals
develop a weak precursor. Then, the neutral jump weakens and disappears.
Finally, the C-shock runs through minor adjustments and reaches the
steady state.

In intermediate Type II shocks (which are quasi-hydrodynamic), the full jump
shock remains with an ion-precursor in advance. Between Type I and Type II,
partial jumps remain. Switch-on shocks develop extremely long ion-magnetic  precursors due to forward-moving undamped ionic Alfv\'en waves. Switch-off
shocks, like all slow shocks, are J-shocks.

We have modelled the time-development of molecular hydrogen emission lines
in the infrared. If the initial jump shock is fast, the molecules do not
survive, and little emission occurs. As the precursor develops, the line
intensities rise rapidly. As the jump shock decreases in strength, the
molecules begin to survive and emit with the high excitation
characteristic of a jump shock (e.g.\ with a 1-0 S(1)/2-1 S(1)
intensity ratio of 3--5). Later, the jump 
is too weak to excite H$_2$ and the C-shock dominates the spectrum,
producing  line ratios that are very sensitive to the particular
parameters of the shock.

The evolution time scale is inversely dependent on the ion
density. For typical molecular cloud parameters, it takes
between (10$^9$/n$_i$)\,s and (2\,10$^{10}$/n$_i$)\,s to set up the
steady-state flow, several times the flow timescale of
(7\,10$^8$/n$_i$)\,s.  Neutralised shocks, modelled here by fixing the
ion 
number density, are approximately M$_a$ times wider than the transverse
equivalent and take the longest to evolve. It follows that detectable changes to the
emission lines occur over periods of at least 60 years for the ion density
taken here (n$_i$ = 0.1\,cm$^{-3}$). However, faster evolving
shocks, due to high ion densities (perhaps in the densest cloud regions), 
will show detectable changes faster.  The time scale for emission-line
variations is $\sim (2\,10^8$/n$_i$)\,s where n$_i$ is the pre-shock
ion number density. 

This study provides the background to  investigations of shock interactions,
C-type jet flows, and multi-dimensional studies such as 
simulations of the Wardle instability (Mac Low \& Smith 1997a,b). We hope to
add additional physics to this version of ZEUS, including chemical
reactions, molecular dissociation and streaming ionisation. 

\vspace*{0.4cm}

{\em Acknowledgments:} MDS thanks the DFG for financial support.

\end{document}